\documentclass[twocolumn,preprintnumbers,showpacs,pre,floatfix,superscriptaddress,amsmath,amssymb]{revtex4-1}
\usepackage{epsfig}
\usepackage{subfigure}
\usepackage{amsmath}
\usepackage{color}
\usepackage{amssymb}
\usepackage{setspace}
\usepackage{graphicx}
\usepackage{dcolumn}
\usepackage{bm}
\usepackage{times}
\input epsf

\begin{document}

\author{Dane Taylor}
\email{dane.taylor@colorado.edu}
\affiliation{Department of Applied Mathematics, University of Colorado, Boulder, Colorado 80309, USA}
\author{Daniel B. Larremore}
\email{larremor@hsph.harvard.edu}
\affiliation{Department of Applied Mathematics, University of Colorado, Boulder, Colorado 80309, USA}
\affiliation{Center for Communicable Disease Dynamics, Department of Epidemiology, Harvard School of Public Health, Boston, Massachusetts 02115, USA}

\date{\today}
\begin{abstract}
Formation and fragmentation of networks is typically studied using percolation theory, but most previous research has been restricted to studying a phase transition in cluster size, examining the emergence of a giant component. This approach does not study the effects of evolving network structure on dynamics that occur at the nodes, such as the synchronization of oscillators and the spread of information, epidemics, and neuronal excitations. We introduce and analyze new link-formation rules, called {\it Social Climber} (SC) attachment, that may be combined with arbitrary percolation models to produce a previously unstudied phase transition using the largest eigenvalue of the network adjacency matrix as the order parameter. This eigenvalue is significant in the analyses of many network-coupled dynamical systems in which it measures the quality of global coupling and is hence a natural measure of connectivity. We highlight the important self-organized properties of SC attachment and discuss implications for controlling dynamics on networks.
\end{abstract}

\title{Social Climber attachment in forming networks produces phase transition in a measure of connectivity}
\maketitle

\section{Introduction}

Dynamics on networks has become a research area of broad importance, with considerable effort focused on understanding how dynamics are affected by network structure \cite{info_diff,restrepo2008-prl,taylor2012-epl,millanese,taylor2011-pre,restrepo2006-prl2,restrepo2005-pre,wang2003,larremore2011-prl,pomerance2009,chakrabarti2008,failure,brain,immunization,perc_theory,sensor,newman}.
Of particular interest are dynamics that depend on global measures of network connectivity, and in particular on the largest eigenvalue $\lambda$ of the network adjacency matrix $A$ ($A_{ij}\not=0$ if a link exists from node $i$ to node $j$). We will refer to this broad class, which includes models for synchronization \cite{restrepo2005-pre}, genetic expression \cite{pomerance2009}, neural excitation \cite{larremore2011-prl}, and epidemic spreading \cite{wang2003,chakrabarti2008}, as {\it connectivity-governed dynamics}. We note, however, that while analyses of such systems \cite{restrepo2005-pre,wang2003,larremore2011-prl,pomerance2009,chakrabarti2008} typically assume that the network structure is static and connected (i.e., lacking isolated nodes/clusters), many applications exist for which the network structure is non-static and/or fragmented, such as epidemic spreading with immunization \cite{immunization,chakrabarti2008}, communication and transit systems operating under failure or attack \cite{failure}, and information processing in the brain \cite{brain}. 

The systems that we categorize as having connectivity-governed dynamics share a common property of being easily manipulated through changes in $\lambda$. For example, one can prevent viral spreading in technological and social networks by decreasing $\lambda$ through immunization \cite{chakrabarti2008} or promote the dissemination of information in communication and sensor networks by increasing $\lambda$ \cite{taylor2011-pre}. In recent years there has been much interest in studying the effects of topological modification on $\lambda$ and developing efficient strategies for tuning $\lambda$ through the addition and/or subtraction of links and/or nodes \cite{taylor2011-pre,millanese,restrepo2006-prl2}. However, such perturbation techniques do not address networks undergoing formation or fragmentation processes, a problem traditionally studied with network percolation theory \cite{perc_theory}. 

From network resilience against targeted attacks and failures \cite{restrepo2008-prl,taylor2012-epl} to dynamics on networks \cite{restrepo2005-pre,wang2003,larremore2011-prl,pomerance2009,chakrabarti2008}, which often map spreading dynamics to percolation processes, there are widespread applications for percolation theory in the field of complex networks. However, applying such techniques to dynamics on evolving networks is hindered in that it can require two levels of analysis, theory for the change in network topology and theory for the dynamics. We hypothesize that a central element frustrating the development of this approach is that this field of research has largely focused on studying a phase transition in the size of the largest connected cluster by analyzing cluster aggregation and the emergence of a {\it giant component}, a cluster whose size is of the same order as the entire system \cite{perc_theory}. However, the application of subsequent percolation theory requires information about cluster topology in addition to cluster size, which highlights the need for percolation theory focusing on other cluster properties such as the spectra of clusters (i.e., the eigenvalues of adjacency matrices corresponding to clusters), modularity, assortativity, transitivity, etc. \cite{newman}.

Here we study the link-percolation phase transition using $\lambda$ as a novel order parameter, shedding light on a new phase transition in connectivity, corresponding to a poorly-connected network becoming well-connected (or vice versa) in terms of the topology's effect on dynamics. In order to produce such a transition, we introduce a link-formation rule called {\it Social Climber} (SC) attachment for which we derive the asymptotic scaling behavior of $\lambda$ for large network size $N$.
We show that networks forming under SC attachment exhibit maximal scaling $\lambda\sim\mathcal{O}(N^{1/2})$, indicating that our model may be of broad interest for the design of networks with large $\lambda$, a property that is often beneficial \cite{taylor2011-pre} and can lead to, for example, excellent robustness against attack and failure \cite{restrepo2008-prl,taylor2012-epl} and very good spreading characteristics \cite{wang2003,larremore2011-prl}.
It follows that SC attachment is a promising approach for the design of self-organized communication and sensor networks \cite{sensor} with topologies designed for the rapid dissemination of information. We demonstrate this application by showing that networks formed under SC attachment exhibit enhanced spreading properties with respect to the susceptible-infected-susceptible (SIS) model \cite{wang2003,chakrabarti2008}, a contagion model with many applications including the dissemination of information, sometimes referred to as ``gossip based'' communication or epidemic routing \cite{info_diff}. We note that the development this potential application may be facilitated by the fact that SC attachment may be combined with arbitrary percolation processes, such as Erd\"os-R\'enyi (ER) percolation \cite{erdosrenyi1959} and Achlioptas processes \cite{AE,exp1,achlioptas2009}, to independently control cluster aggregation (determined by the percolation process) and connectivity within clusters (determined by SC attachment). 

\section{Social climber attachment}

\begin{figure}[t]
	\includegraphics[width=1.0\linewidth]{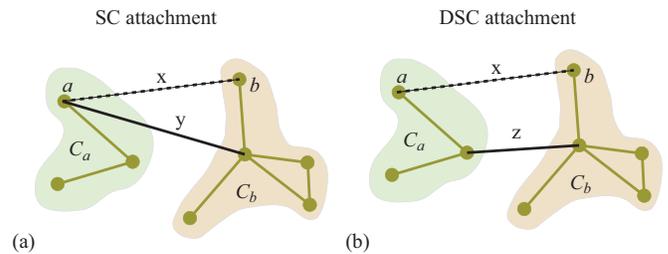} 
	\caption{(color online) Link $x$ is proposed by an arbitrary percolation model (dashed lines) to connect nodes $a$ and $b$, merging clusters $C_{a}$ and $C_{b}$. Because $C_{a} \neq C_{b}$ the proposed link $x$ is discarded and instead one of the following new links is formed. 
(a) Node $a$ is linked to the largest-degree node in $C_{b}$ with link $y$ to model SC attachment.
(b) The largest-degree nodes from $C_{a}$ and $C_{b}$ are linked together with link $z$ to model DSC attachment. These processes may be visualized using free PercoVIS software \cite{percovis}.}
	\label{SCcartoon}
\end{figure}

A {\it link-percolation process} begins with $N$ isolated nodes, indexed $n=1,2,\dots,N.$ In discrete steps $\hat{t} = 1,2,\dots$, a new undirected link between two nodes is selected according to a rule or set of rules, and is then formed. Thus, after $\hat{t}$ steps there will be $\hat{t}$ links in the network, resulting in clusters of connected nodes, each of whose size (number of nodes in the cluster) may range from one (an isolated node) to $N$ (a cluster that spans the entire network). Depending on the rules used to select links, the evolution of cluster sizes, and in particular the size of the largest cluster, may vary significantly. Social Climber attachment introduces a new link reselection step between link selection and link formation, which we motivate by analogy to a corresponding social process: colloquially, a ``social climber'' is someone who actively attempts to make powerful friends in order to become more powerful himself. When introduced to a new person, a social climber learns about the relative popularity of the people in that person's clique and eventually befriends whoever is of maximal importance.
With this in mind, SC attachment is a link reselection step during percolation where the proposed link between two nodes is altered by allowing one of those nodes to act like a social climber, choosing to link to the node of maximal importance in the other node's cluster. Therefore, given a link-percolation process, we summarize SC attachment as follows.
(i) Let $x$ be a proposed undirected link connecting nodes $a$ and $b$, generated by an arbitrary percolation model. %
(ii) Let clusters $C_a$ and $C_b$ be the clusters to which nodes $a$ and $b$ belong, respectively. Then, if $C_a\not=C_b$ the proposed link $x$ is discarded and instead a link $y$ is made between node $a$ and the largest-degree node in $C_b$, as shown in Fig.~\ref{SCcartoon}a.
(iii) If nodes $a$ and $b$ belong to the same cluster, $C_a=C_b$, then the proposed link $x$ is made without modification. 
Note that SC attachment does not affect which clusters combine, but does affect the topology of the resulting joined cluster.
The SC model chooses a connection to the node of largest degree in a cluster using nodal degree as a proxy for the {\it dynamical importance} measure, $DI= u_{n}v_{n}$ \cite{restrepo2006-prl2}, where $u$ and $v$ denote the right and left eigenvectors of $A$ corresponding to $\lambda$. For the undirected networks considered here, symmetry of $A$ implies $u=v$, and thus the node with largest eigenvector entry $u_{n}$ will have maximal $DI$ in its cluster \cite{restrepo2006-prl2}. Provided that the node with largest eigenvector entry $u_{n}$ also has largest degree $k_{n}$, we allow the SC model to select nodes based on degree for simplicity and ease of computation. One may equivalently view SC attachment as forming a link to the node with largest degree by using degree centrality as a proxy for eigenvector centrality.

In addition to SC attachment, we introduce {\it Double Social Climber} (DSC) attachment, in which a proposed link between nodes $a$ and $b$ is either replaced by a link between the nodes with maximal degree in each cluster when $C_{a} \neq C_{b}$, as shown in Fig. \ref{SCcartoon}b, or is formed between $a$ and $b$ without modification when $C_{a} = C_{b}$. We note that DSC attachment corresponds to maximizing connectivity of the resulting cluster, as measured by $\lambda$, whenever the node of maximal degree is also the node of maximal eigenvector entry in each cluster \cite{taylor2011-pre}. SC and DSC attachment may be visualized using free PercoVIS software \cite{percovis}. 

\section{Analysis}

Although we will later generalize our methods to other percolation rules, we first analyze SC and DSC attachment for the well-known ER percolation process \cite{erdosrenyi1959}. The rule for selecting a link in ER percolation is simple: two nodes are chosen uniformly at random and a link is formed if there is not already a link between them.
Traditional analysis has focused on the relationship between the number of links added $\hat t$ and the size of the largest cluster $\hat G ( \hat t )$, called the {\it giant component} (GC) when $\hat{G}(\hat{t})\sim\mathcal{O}(N)$. It is convention to rescale both $\hat{t}$ and $\hat{G}$ by $N$ [i.e. $t = \hat{t}/N$ and $G(t) = \hat{G}(t)/N$], where one obtains in the asymptotic limit $N\to\infty$ \cite{erdosrenyi1959},
\begin{equation}
	G(t) = \left \{ \begin{array}{ccc} 
		0 &,&t\le 0.5 \\
		1 - e^{-2tG(t)} &,&t> 0.5
	\end{array} \right. .
	\label{ER_eq}
\end{equation} 
Here, for variable control parameter $t$, the network undergoes a second order phase transition in cluster size at the percolation threshold $t^{ER}_c=0.5$, as observed through the order parameter $G(t)$.
Because SC attachment affects the topology of clusters and not their sizes, Eq.~\eqref{ER_eq} remains valid for ER percolation combined with SC attachment. 

We begin our analysis by studying the emergence of large-degree nodes. For a given time $t$, consider a large cluster $C$ containing $s\gg1$ nodes, and let $k^{\text{max}}$ denote the maximal nodal degree in $C$. By large cluster, we mean that $s$ is with high probability larger than the size of another randomly chosen cluster in the network, and eventually we will consider only the case in which cluster $C$ is the largest cluster in the entire network. We will compute the expected change in $k^{\text{max}}$ for the addition of a single link. When a link is proposed between nodes $a$ and $b$ by ER percolation, $k^{\text{max}}$ will increase by one if: (i) $a \notin C$ and $b \in C$ (depicted in Fig.~\ref{SCcartoon}a where $C_b=C$), or (ii) $a,b \in C$ and the degree of $a$ or $b$ is $k^{\text{max}}$. Since  ER percolation chooses nodes uniformly at random, the probability that $a \notin C$ is $1-s/N$, and the probability that $b \in C$ is $s/N$. Since $a$ and $b$ are chosen independently, the probability of case (i) is $(s/N)[1-s/N]$. The probability that a randomly chosen node in a cluster of size $s$ has degree $k^{\text{max}}$ is $r/s$, where $r$ is the number of nodes in that cluster with degree $k^{\text{max}}$. Thus, the probability of case (ii) is, to leading order as $N \to \infty$, $(s/N)[1-s/N](r/s) + (s/N)^2[2r/s]$, where the first term corresponds to $a \in C$, $b \notin C$ and the second term corresponds to $a,b \in C$. We note that other corrections may be included to address the chance in (i) that the maximal degree of $C_{a}$ is larger than the maximal degree of $C_{b}$, $k^{max}+1$, but such corrections decay rapidly as the size difference between $C_{b}$ and $C_{a}$ increases.  
Since $s\gg1$ while $r\sim\mathcal{O}(1)$, case (i) is the dominating process for sufficiently large $s$, so the expected rate of change of $k^{\text{max}}$ averaged over all possible links is
\begin{equation}
	\label{maxdegreegrowthrate} 
	E\left [\frac{d k^{\text{max}} }{d\hat{t}} \right] = \zeta {\frac{s}{N}\left(1-\frac{s}{N} \right)}.
\end{equation}
Here, $\zeta=1$ for SC attachment and $\zeta = 2$ for DSC attachment, since for DSC case (i) applies to both $a \notin C$, $b\in C$ and $a \in C$, $b\notin C$.
Using $d/d\hat{t} = Nd/dt$, integration of Eq.~\eqref{maxdegreegrowthrate} predicts that the largest degree of a node within the GC at time $t$ is 
\begin{equation}
	\label{maxdegreegc}
	E\left[ k_{ER}^{\text{max}}(t) \right]= \zeta_{}N \int_0^t \left[G(\tau) - G^2(\tau) \right]d\tau.
\end{equation}
Note that this scaling with $N$ is the largest achievable scaling of a degree.
For comparison, in networks with power-law degree distribution, $P(k)\varpropto k^{-\gamma}$, the expected maximal degree scales as $\mathcal{O}(N^{1/(\gamma-1)})$, approaching $\mathcal{O}(N)$ as $\gamma\to 2^+$. 

\begin{figure}[t]
	\includegraphics[width=1.0\linewidth,trim=3mm 1mm 10mm 1mm,clip]{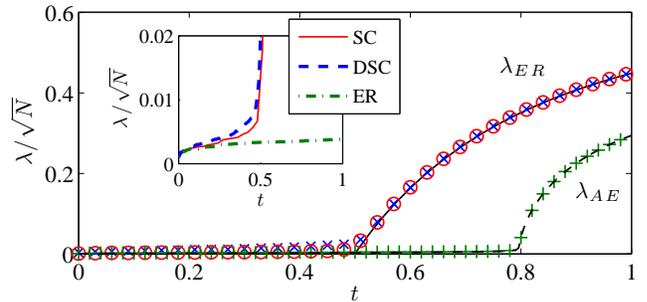}
	\caption{(color online) $\lambda_{ER}$ is shown for ER percolation with DSC attachment. Our prediction, Eq.~\eqref{eval} with $G(t)$ given by Eq.~\eqref{ER_eq} (solid line), agrees well with observed values for simulations with $N=10^{5}$ (X symbols) and $N=10^{6}$ (circles). Observed values for $\lambda_{AE}$ (crosses) also agree well with theory, Eq.~\eqref{eval_PR} (dashed line), for AE percolation with DSC attachment for $N=10^6$. (inset) For $N=10^6$, the great extent to which SC and DSC attachment increase connectivity is shown by comparing to $\lambda$ for classical ER percolation. }
	\label{fig:evalue}
\end{figure}

In order to understand the implications of Eq.~\eqref{maxdegreegc}, we use $\lambda\approx\sqrt{k^{\text{max}} }$, an asymptotically ($N\to \infty$) accurate approximation derived in \cite{chung2003} and discussed further in \cite{restrepo2007}. While the model used to generate this estimate is not equivalent to SC attachment, we find it remains accurate here.
Using this estimate in conjunction with Eq.~\eqref{maxdegreegc}, and noting that for $t > t^{ER}_{c}$, the largest eigenvalue of the GC will be larger than the largest eigenvalues of smaller clusters, we obtain the following expression for the expected largest eigenvalue for ER percolation with SC attachment,
\begin{equation}
	\displaystyle E\left[\lambda_{ER}(t)\right]  =\sqrt{\zeta N \int_0^t \left[G(\tau) - G^2(\tau) \right]d\tau},
	\label{eval}
\end{equation}
implying that the network undergoes a continuous phase transition in connectivity at precisely the same value $t=t_c^{ER}$ at which a phase transition in cluster size occurs. In the super-critical regime, $\lambda$ achieves maximal scaling, $\lambda \sim \mathcal{O}(N^{1/2})$. For comparison an all-to-all network has similar scaling, $\lambda = \sqrt{N-1}$, but uses $\mathcal{O}(N^{2})$ links compared to $\mathcal{O}(N)$ used by SC attachment.
Asymptotic scaling constants of Eq.~\eqref{eval} for large $t$ may be solved by integrating with respect to $G$, rather than $t$, and using a dilogarithm to obtain $\lambda_{ER}(t)/\sqrt{N} \to \sqrt{1-\pi^2/12}\approx0.42$ for SC attachment and $\lambda_{ER}(t)/\sqrt{N} \to \sqrt{2-\pi^2/6}\approx0.6$ for DSC attachment \cite{note1}.
We confirm the accuracy of Eq.~\eqref{eval} by direct simulation of our model, shown in Fig.~\ref{fig:evalue}, which demonstrates excellent agreement between Eq.~\eqref{eval} (solid line) and observed values of $\lambda_{ER}$ for ER percolation with DSC attachment with $N=10^{5}$ (X symbols) and $N=10^{6}$ (circles). For integration in Eq.~\eqref{eval} we use the asymptotic theoretical value $G(t)$ given by Eq.~\eqref{ER_eq}. In the inset of Fig.~\ref{fig:evalue} we compare observed values of $\lambda_{ER}$ for SC and DSC attachment to traditional ER percolation, where our models' main effects are highlighted: under SC or DSC attachment $\lambda_{ER}$ attains significantly larger values and also undergoes a sharp increase at $t_{c}^{ER}$. 

\begin{figure}[t]
	\includegraphics[width=1.0\linewidth,trim=5mm 1mm 15mm 1mm,clip]{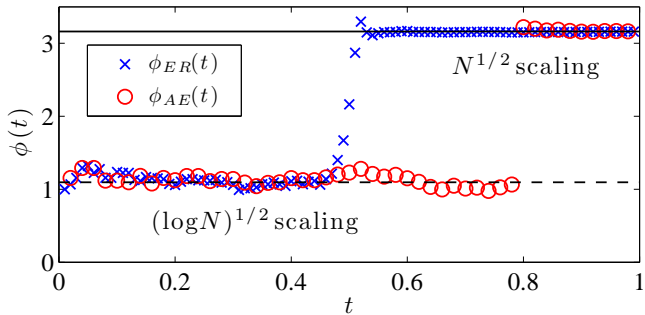}
	\caption{(color online) Scaling of $\lambda_{ER}(t)$ predicted in Eq.~\eqref{scaling} is demonstrated using the ratios of $\lambda_{ER}(t)$ for $N=10^6$ and $N=10^5$, denoted $\phi_{ER}(t)$. Agreement between the prediction of Eq.~\eqref{scaling} (lines) and measurement from simulation (crosses) is good. Measurements from AE percolation, $\phi_{AE}(t)$, are also shown (circles).}
	\label{fig:scaling}
\end{figure}

Because the expected connectivity, measured by $\lambda_{ER}$, is a function of $G(\cdot)$, developing scaling arguments for $\lambda_{ER}$ is straightforward since scaling for $\hat G(t)$ is known: $\hat{G}\sim\log(N)$ for $t<t^{ER}_c$ 
and $\hat{G}\sim N$ for $t>t^{ER}_c$ \cite{erdosrenyi1959}. Therefore, when SC is used in conjunction with ER percolation, in the limit $N \to \infty$ we have
\begin{align}
	\lambda_{ER}(t) &\sim \sqrt{\log N} &t < t^{ER}_c, 
	\notag
	\\
	\lambda_{ER}(t) &\sim N^{1/2} &t > t^{ER}_c.
	\label{scaling}
\end{align}
To validate Eq.~\eqref{scaling} we estimate the change in $\lambda_{ER}$ when system size is increased by defining $\phi_{ER}(t)$ as the ratio of $\lambda_{ER}(t)$ for $N=10^6$ to $\lambda_{ER}(t)$ for $N=10^5$. As shown in Fig.~\ref{fig:scaling}, we predict that $\phi_{ER}(t) \approx \sqrt{6/5}$ for $t < t^{ER}_{c}$ (dashed line) and $\phi_{ER}(t) \approx \sqrt{10}$ for $t > t^{ER}_{c}$ (solid line), both of which agree well with $\phi_{ER}(t)$ calculated from a single simulation of each system size (X symbols).

The methods used to derive Eqs.~(\ref{maxdegreegc}-\ref{scaling}), which involved calculating the probability that an isolated cluster attaches to the GC, may be easily adapted to other percolation models. For example, consider Achlioptas processes \cite{AE,exp1,achlioptas2009} for which the merging of clusters depends on cluster size (up to some bound). This class of percolation models has recently received much attention, focusing on analysis of a rapid phase transition in cluster size referred to as ``explosive percolation'' \cite{AE,exp1,explosive}. Repeating the reasoning process in deriving Eqs.~(\ref{maxdegreegc}-\ref{scaling}) for Adjacent-Edge (AE) percolation \cite{AE} , we predict the largest eigenvalue to be
\begin{equation}
	E[\lambda_{AE}(t)]  =\sqrt{\frac{\zeta N}{2} \int_0^t  \left[G(\tau)+ G^2(\tau)-2G^3(\tau) \right]d\tau},
	\label{eval_PR}
\end{equation}
where again $\zeta=1$ for SC and $\zeta=2$ for DSC. Note that despite the near-discontinuous phase transition in $G$, maximal scaling $\lambda\sim\mathcal{O}(N^{1/2})$ is still achieved. 
In Fig.~\ref{fig:evalue} we show good agreement between observed values for $\lambda_{AE}$ (crosses) and Eq.~\eqref{eval_PR} for DSC (dashed line), where observed values for $G(t)$ were used in Eq.~\eqref{eval_PR} as an analytic expression has yet to be developed. 
Note that $\lambda_{AE}(t) < \lambda_{ER}(t)$, which we attribute to the integrands of Eqs.~\eqref{eval} and \eqref{eval_PR}, which are maximized at $G=1/2$ and $G=(1+\sqrt{7})/6$ respectively and are zero at $G=0$ and $G=1$. Since AE percolation produces rapid growth in $G$, the integrand of Eq.~\eqref{eval_PR} is not large over a majority of the integration interval, so essentially, the explosive growth in $G$ minimizes the regime during which SC attachment has a large effect on $\lambda$. 
In Fig.~\ref{fig:scaling} we also show $\phi_{AE}(t)$, the ratio of $\lambda_{AE}(t)$ for $N=10^6$ and $N=10^5$, where we observe similar scaling in the subcritical and supercritical regimes as observed and predicted for $\lambda_{ER}$. 

\section{Experimentation}

We now demonstrate the effect of SC attachment on dynamics. Because SC attachment produces networks with maximal scaling of $\lambda$, we focus on an application in which large $\lambda$ is beneficial: the dissemination of information in communication and wireless sensor networks \cite{sensor}, which is often modeled as an epidemic \cite{info_diff}. We note, however, that large $\lambda$ is not always advantageous. For example, large $\lambda$ in ecological networks can promote instability and species extinction \cite{eco}. (See \cite{taylor2011-pre} for a discussion of applications in which it is beneficial to have either small or large $\lambda$.) 
Here we study SIS contagion \cite{wang2003,chakrabarti2008}, which has been used to study spreading process from viral propagation in social and technological networks to the dissemination of information such as rumors and data \cite{info_diff}. 

To briefly review, the SIS model is a continuous time process in which each node may be {\it susceptible} to infection or {\it infected}. Each infected node may infect each of its susceptible network neighbors at rate $\alpha$, and each infected node may also spontaneously heal and return to being susceptible at rate $\beta$. The network state in which no nodes are infected and all nodes are susceptible is a fixed point of the collective dynamics, but this fixed point may not be stable to perturbation (i.e., a small fraction of nodes being infected by some external agent). For many topologies of connected networks in which a fraction of nodes are initially infected, the expected steady-state fraction of infected nodes $f$ may either be zero (no infections, stable fixed point) or nonzero (endemic infection, unstable fixed point), depending on whether $\alpha/\beta$ surpasses the epidemic threshold $\lambda^{-1}$ \cite{wang2003,chakrabarti2008}. 
Note that endemic infection can be prevented by decreasing $\lambda$ through immunization until $\lambda^{-1}>\alpha/\beta$. Interestingly, for very large infection rates, reducing $\lambda$  to prevent endemic infection can require the complete fragmentation of the network. For example, if $\alpha/\beta \geq 0.5$ the prevention of endemic infection requires $\lambda \leq 2$ which guarantees fragmentation of the network \cite{taylor2012-epl}.
This scenario has been observed experimentally for virus propagation on mobile phone devices \cite{wang2009}, where slowly spreading Bluetooth viruses may be inhibited by immunization (i.e., antiviral software) but rapidly spreading messaging viruses are inhibited only by a fragmented network.

\begin{figure}[t]
	\includegraphics[width=1\linewidth,trim=5mm 1mm 10mm 1mm,clip]{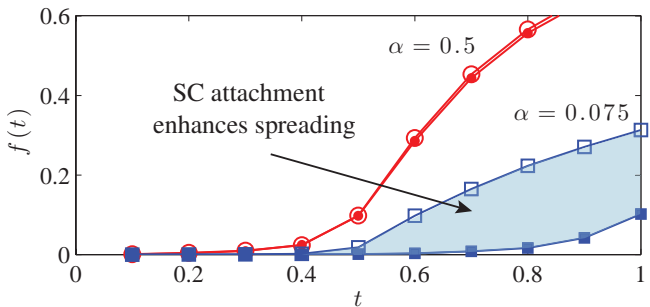} 
	\caption{ (color online) SIS epidemics \cite{wang2003} were simulated on a network forming by ER percolation with SC attachment (open symbols) or without (filled symbols), where $tN$ is the number of links added. $f(t)$ is the steady-state fraction of infected nodes when 1\% of nodes are initially infected. The shaded region highlights the significant impact of SC attachment for moderate infection rate (squares). For high infection rate, SC attachment has no effect (circles).}
	\label{fig:SIS}
\end{figure}

We simulated SIS dynamics for moderate $\alpha/\beta$ on two networks forming under ER percolation, one with SC attachment and the other without, predicting that SC attachment will have a significant impact on the steady state fraction of infected nodes, $f$. We simulated dynamics with $(\alpha,\beta)=(0.075,1)$ on $N=10^5$ nodes at many points $t$ in the percolation process, initially infecting 1\% of nodes and then allowing the system to reach a steady state fraction of infected nodes $f(t)$, before allowing percolation to continue to another value of $t$, where the dynamics were re-initialized, simulated, and so on. The resulting curves $f(t)$, are shown in Fig.~\ref{fig:SIS}, where the shaded region highlights that networks forming with SC attachment (open squares) have significantly enhanced spreading characteristics compared with networks forming without SC attachment (filled squares). 
To contrast this result, we also plot $f(t)$ for a large infection rate $(\alpha,\beta)=(0.5,1)$, where $f(t)$ with SC attachment (open circles) is indistinguishable from that without SC attachment (filled circles). This is not surprising as one would expect any initial infection to saturate the cluster in which it begins, in which case $f(t)$ would depend primarily on cluster size, not topology. 
We thus find two regimes of SIS dynamics on fragmented networks: when $\alpha/\beta$ is sufficiently large, $f$ depends primarily on network fragmentation (i.e. the size of the GC), but for moderate and small $\alpha/\beta$, $f$ depends strongly on the connectivity of clusters (i.e., their respective $\lambda$ values). 

\section{Discussion}

Motivated by the need for the development of analysis for connectivity-governed dynamics \cite{restrepo2005-pre,wang2003,larremore2011-prl,pomerance2009,chakrabarti2008} on evolving networks, we have developed a percolation theory focusing on the connectivity of clusters, rather than their size.
In this pursuit, we have introduced a model, Social Climber attachment, that produces networks with strong connectivity and maximal scaling of $\lambda$, and validated our claims using two link-percolation models. While strong connectivity in networks is achievable via other percolation models (e.g., networks with heavy-tailed degree distributions generated by the Chung-Lu model \cite{chung2002}), such methods typically require that the nodal degrees and network connectivity be defined {\it a priori}. In contrast, two key properties distinguish SC attachment. 
(i) First, SC attachment produces networks with large $\lambda$ via self-organization. Because $\lambda$ governs many dynamical processes \cite{restrepo2005-pre,wang2003,chakrabarti2008,larremore2011-prl,pomerance2009}, SC attachment provides a foundation for designing networks that self-organize with properties linked to large $\lambda$ such as robustness \cite{restrepo2008-prl,taylor2012-epl} and the efficient spread of information \cite{taylor2011-pre}. Our model is therefore promising as a starting point for the development of self-organized communication networks such as wireless sensor networks \cite{sensor}, where data broadcasting may be modeled by SIS transmission \cite{info_diff}. Development and analysis of this application may be facilitated by the fact that SC attachment does not affect cluster sizes, only their internal topology.
(ii) Second, a novel phase transition in connectivity occurs for networks forming under SC attachment, which may have broad applications. For example, $\lambda(t)$ changes most rapidly near the percolation threshold, so creating networks near criticality may offer an effective approach for designing networks on which dynamics can be efficiently controlled by adding or removing a minimal number of links. 
This approach may therefore aid in the design of critical infrastructure (e.g., the power grid, communication networks, and airline networks) that can be easily switched between topologies designed for high-flow and low-flow conditions.

 We conclude by suggesting several possible extensions to this work that may be of interest to readers. 
First, SC attachment uses complete information about the structures of the clusters that it connects, yet in some applications this information may be difficult or impossible to obtain. The effects of incomplete information or noise on SC attachment are as yet unexplored. For example, incorporating a probabilistic (rather than deterministic) link reselection step may be of interest as the resulting process would have some similarity with the preferential attachment network growth model \cite{Price}.
Second, one may wish to adapt our model to study various real-world networks. While networks formed under SC attachment feature large $\lambda$, to incorporate this model for the design of communication networks one would likely need to consider many other design criteria such as betweenness centrality and software protocols \cite{sensor}. Finally, we named our model Social Climber attachment to reflect the selfish behavior of individuals in social situations, yet the generation of a network topology similar to that observed in social networks using the SC model would require additional link-formation rules, such as those producing modularity and transitivity \cite{newman}.

\acknowledgements
We thank R. S. Maier for suggesting the transformation of Eq.~\eqref{eval} to a dilogarithm, J. G. Restrepo and P. S. Skardal for many insightful discussions, and the referees for their helpful suggestions. 
Funding for D. T. was provided by NSF Grant No. DMS-0908221. 
Funding for D. B. L. was provided by NSF MCTP Grant No DMS-0602284.

\bibliographystyle{plain}

\begin{thebibliography}{99}
\bibitem{note1} Such approximations are accurate for large $t$ but once $t \sim \mathcal{O}(N^{1/2})$, SC or DSC attachment will no longer be the dominant process governing the growth of $\lambda$, since for ER percolation without SC attachment, $\lambda(t) \approx 2t$.
\end{thebibliography}

\end{document}